\documentstyle[12pt,epsf]{article}

\def\Journal#1#2#3#4{{#1} {\bf #2}, #3 (#4)}

\def\NPB{{\em Nucl. Phys.} B}
\def\PLB{{\em Phys. Lett.}  B}
\def\PRL{\em Phys. Rev. Lett.}
\def\PRD{{\em Phys. Rev.} D}
\def\ZPC{{\em Z. Phys.} C}

\def\be{\begin{equation}}
\def\ee{\end{equation}}
\def\bea{\begin{eqnarray}}
\def\eea{\end{eqnarray}}

\begin{document}

\begin{flushright}
ITP-SB-96-62\\
October, 1996\\
hep-ph/9610462
\end{flushright}
\vbox{\vskip 0.75 true in}

\centerline{\large \bf HARD SCATTERING IN QCD\footnote{Presented at the
28th International Conference on  High Energy Physics,
July 1996, Warsaw, Poland.}}
\vbox{\vskip 0.25 true in}

\centerline{\large  GEORGE STERMAN}
\vbox{\vskip 0.25 true in}

\centerline{Institute for Theoretical Physics} 
\centerline{State University of New York at Stony Brook}
\centerline{Stony Brook, NY 11794-3840, USA}
\vbox{\vskip 2.0 true in}

\begin{abstract}
Recent developments in the theory and experiment of QCD hard scattering
are described.
\end{abstract}

\section{Preamble:  How Perturbative QCD Works}

This has been an active year for experimental results and theoretical
developments in quantum chromodynamics.  In this talk I will cover
a selection of topics presented at this conference, tied together by
the interplay of soft and hard physics.  QCD is certainly the
most complex component of the standard model at current energy
scales, evolving as it does from an asymptotically-free perturbative
theory of quarks and gluons at short distances to a confining
theory of hadrons at long distances.  
As experiment and theory become more precise,
it becomes both possible and necessary to explore the relations
between these two limits.  I begin with a short review of 
how we employ the simplicity of the theory at short distances even
in the presence of nonperturbative low-energy dynamics.
Thus, let us recall a few recurring themes in QCD theory. \cite{gstasi}

{\em Infrared Safety}.  Our first observation is that
there is a class of high-energy cross sections and decay rates that may be calculated
directly in perturbation theory.  These ``infrared safe" quantities are
free of long-distance dependence at leading {\it power} in a large momentum
scale $Q$, such as the total c.m.\ energy for ${\rm e}^+{\rm e}^-$ annihilation
(at lowest order in electroweak couplings).  Such a quantity, $\hat{\sigma}$
may be written as a power series in the strong coupling evaluated
at scale $\mu$,
\be
Q^2\hat{\sigma}(Q^2,\mu^2,\alpha_s(\mu^2))=\sum_n c_n(Q^2/\mu^2)\alpha_s^n(\mu^2)\, ,
\label{irs}
\ee
where the $c_n$ are infrared finite coefficients, free of dependence on
long-distance parameters.  
Corrections due to dimensional parameters such as quark masses and  
 vacuum condensates are suppressed by powers of $Q$.  Other examples are
jet cross sections in  ${\rm e}^+{\rm e}^-$ annihilation and the total width of
the Z boson.   

{\em Factorization}. The class of infrared-safe cross sections is a small
subset of all high-energy cross sections.  In particular,
hard-scattering cross sections
that involve one or  more hadron in the initial state cannot be infrared safe.
This is
because they depend on the distributions of quark and  gluon degrees of
freedom in the hadrons, which reflect the nonperturbative, long-time processes that
bind the hadrons.  Nevertheless, it is often possible
to factorize such a cross section into a  convolution of an
infrared-safe short-distance
``hard part" $\hat{\sigma}$, calculable in perturbation theory
and specific to the process, and a long-distance
function $\phi_{\rm NP}$, which, although nonperturbative, is universal
among different processes. Thus, for example, a deeply inelastic
scattering (DIS) cross section may be written as
\bea
Q^2\sigma_{\rm phys}(Q,m)&=&\hat{\sigma}_{\rm PT}(Q/\mu,\alpha_s(\mu))
\nonumber \\
&\; & \quad \quad \otimes\;  \phi_{\rm NP}(\mu,m)\, ,
\label{factotimes}
\eea
where the subscript ``phys" on the left emphasizes that this is the physically observed
cross section. The parameter $\mu$ on the right separates short-distance dynamics,
incorporated into $\hat{\sigma}_{\rm PT}$, and long-distance dynamics, at the scale
of parameters $m$,
absorbed into $\phi_{\rm NP}$.  For DIS, the convolution is in terms of the familiar
fractional momentum $\xi$ of the parton that initiates the scattering, and
implicitly includes a sum over parton types.
Factorizations of this sort are at the basis of much of perturbative QCD (pQCD).
The same physical principles recur in other field-theoretic contexts, under
different names, such as the identification of effective field theories.

{\em Evolution}.
Perhaps the most powerful result in pQCD is evolution with momentum transfer.
This is not a single theorem, but a wide-ranging method, which 
can be exploited whenever we succeed in factorizing dynamics at different
length scales, as above.  
The essential observation is that the scale $\mu$ in Eq.\ \ref{factotimes}
is arbitrary, subject only to $\alpha_s(\mu)\ll 1$ (so that $\hat{\sigma}$
may be truly perturbative).
The {\it physical} cross section, however, must be independent
of the choice of this scale of separation,
\be
\mu{d\sigma_{\rm phys}\over d\mu}=0\, .
\label{sigphys}
\ee
The chain rule then leads to separate equations for the
short- and long-distance components of the fields,
\bea
\mu{d\phi_{\rm NP}\over d\mu}&=&P(\alpha_s(\mu))\otimes\phi_{\rm NP}+{\cal O}(1/Q^2)\nonumber \\
\mu{d\hat{\sigma}_{\rm PT}\over d\mu}
&=&-P(\alpha_s(\mu))\otimes\hat{\sigma}_{\rm PT}+{\cal O}(1/Q^2)\, ,
\label{dglap}
\eea
where the convolution is of the same form as in Eq.\ \ref{factotimes}.
Applied to QCD, these are the DGLAP evolution equations. \cite{dglap}
The ``evolution kernel" $P$ plays the role of a parameter
in the separation of variables between the factors $\hat{\sigma}$ 
and $\phi$.  
It can depend on $\alpha_s(\mu)$ and momentum fractions only,
because these are  the only variables that  
$\hat{\sigma}$ 
and $\phi$ hold in common.  Thus, the perturbative
calculability of the familiar DGLAP evolution kernels
is a direct result of the factorizability of cross sections.

\section{From Low $x$ to Reggeons via BFKL}

With this perspective on the basic methods of QCD, we are ready
to turn to one of the major topics of this conference, the
new results on low-$x$ structure functions. \cite{habr}  This will lead us naturally
into a discussion of the theory of ``small-$x$" and
its relation to an old friend, the Regge limit.

\subsection{The ZEUS, H1 1994 Data}

\begin{figure}
\centerline{\epsffile{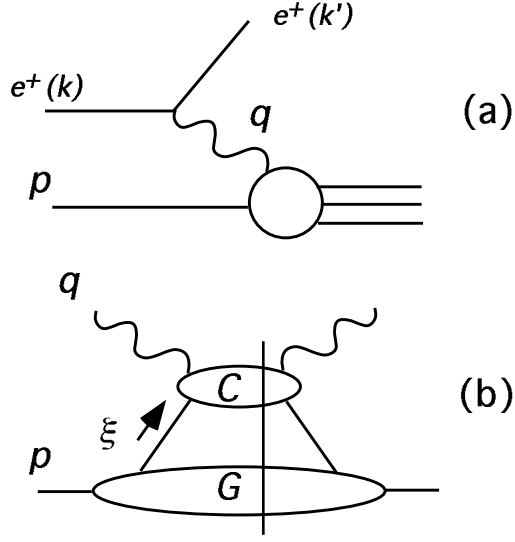}}
\caption{(a) Deeply inelastic scattering.  (b) DGLAP factorization
of a DIS structure function.}
\label{DIS}
\end{figure}

At the colliding beams of HERA, lepton-proton deeply inelastic scattering has been brought
to a new kinematic range, with an unprecedented sensitivity to soft partons. \cite{shekH1,LuZEUS}
These results are often presented in terms of the dimensionless structure function $F_2$,
which for electromagnetic scattering (Fig.\ \ref{DIS}a) is related to the
cross section at fixed momentum transfer $q^2=-Q^2$ and variables $x=Q^2/2p\cdot q$
and $y=p\cdot q/p\cdot k$,
with 
$k$ and $p$ the positron and proton momenta, respectively.
To a good approximation we have, with $s=(p+k)^2$,
\be
{d\sigma\over dxdy}={2\pi\alpha^2\big(1+(1-y)^2\big)s\over Q^4}F_2(x,Q^2)\, .
\label{dsigdxdqsq}
\ee
Data from the 1994 run of H1 and ZEUS cover the general range 
of $10^{-6}<x<10^{-2}$ for momentum transfers $0.3<Q^2<4$ GeV,
over which pQCD ``turns on".  The data for  $10^{-4}<x$ extend
up to much  higher $Q^2$.  We recall that, as in Eq.\ \ref{factotimes},
to lowest order in $\alpha_s$, $x$ has the interpretation of the
fractional momentum of the struck parton.  The range in
$Q^2$ is limited at the smallest values of $x$ by the
kinematic restriction that the total c.m.\ energy 
must be larger than the invariant mass of the hadrons
in the  final  state,
\be
(p+k)^2\ge W^2={Q^2\over x}(1-x)\, .
\ee
The data \cite{habr,shekH1,LuZEUS} shows a distinct rise in $F_2$ toward $x=0$, which
becomes increasingly striking as $Q^2$ increases.   
In the range of $x\sim 10^{-6}$, $xG(x)$ increases to the
order of 30.  To put this value into perspective, we may
recall a typical scaling parton distribution, consistent with early
DIS experiments, $xG(x)=3(1-x)^5$.   

Our current understanding of this rise is based on the evolution
equations for nonsinglet ($f_{\rm ns}$) and singlet 
($f^{\rm (s)}_q=\sum_f(q_f(x)+\bar{q}_f(x)), f^{\rm (s)}_G=G(x)$)
sectors of nucleon parton distributions,
\bea
{dq_{\rm ns}\over d\ln Q^2}&=&P_{qq}\otimes q_{\rm ns}\nonumber \\
{d f^{\rm (s)}_i\over d\ln Q^2}&=&P_{ij}\otimes f^{\rm (s)}_j\, ,
\label{evol}
\eea
which are the practical versions of the general 
relations (\ref{dglap}) above, for quarks and gluons.  
In the singlet (matrix) equation, the gluon-gluon
kernel is singular as $x$ vanishes,
\be
P_{GG}(x)={\alpha_s\over \pi}C_A{1\over x}+\dots\, ,
\label{pxinv}
\ee
where $C_A=N=3$ for QCD.  This singularity, the only one of its
kind in eq.\ \ref{evol}, is generally recognized as the 
origin of the rise in $F_2$ for small $x$.  The rise is
not unexpected, but its {\it detailed} origin is
still debated. \cite{H1analy,Yn}   Most of us agree that it is
the gluon density that is growing for small $x$, but there are  at least two  
possibilities for the dynamics of its growth.  
They may be illustrated by a simplified evolution
equation, involving only the gluonic distribution,
and the singular term in $P_{GG}$ from Eq.\ \ref{pxinv},
\bea
{d\; xG(x,Q^2)\over d\ln Q^2}&=& x\int_x^1{d\xi\over \xi}\Bigg ({N\alpha_s(Q^2)\over \pi(x/\xi)}\Bigg )
G(\xi,Q^2) \nonumber \\
&=&N{\alpha_s(Q^2)\over\pi}\int_x^1 {d\xi\over \xi}\big [\xi G(\xi,Q^2) \big]\, .\nonumber\\
\label{Gvol}
\eea
We consider two possibilities for the gluon distribution
at the value $Q_0^2$ at which we begin the evolution: either (i)
$\xi G(\xi,Q_0^2)\sim 1$, or (ii) $\xi G(\xi,Q_0^2)\sim \xi^{-\lambda}$.
In the first case the $\xi$ integral in Eq.\ \ref{Gvol} 
is itself logarithmic in $x$, $\int_x(d\xi/\xi)\sim \ln x$,
and the $x$-dependence of the distribution changes, and steepens,
as $Q^2$ increases.  To leading logarithm in $x$ and $Q$, case
(i) results in a well-known $x$-dependence that reflects the nature of
the kernel, 
\be
xG(x,Q^2) \sim xG(x,Q_0^2){\rm e}^{4\sqrt{(N/b_2)\ln(1/x)\ln(t)}}\, ,
\label{kerneldrive}
\ee
where $t\equiv \ln(Q^2/\Lambda^2)/\ln(Q_0^2/\Lambda^2)$, with $\Lambda$
the scale of the QCD running coupling, and $b_2=11-2n_f/3$ 
the one-loop coefficient from the QCD beta function.  Notice that this $x$-dependence,
although more singular than $1/x$, and increasingly steep with
$Q$, remains still less singular than any power like $1/x^{1+\lambda}$.

To this ``kernel-driven" $x$-dependence
 we may contrast case (ii), for
which the $\xi$ integral in the evolution equation leaves
the $x$-dependence stable, since then
\be
{d\; xG(x,Q^2)\over d\ln Q^2} \sim x^{-\lambda}\, .
\label{powerstable}
\ee
In summary, if the initial conditions for $\xi G(\xi,Q^2)$ are
smooth as $\xi\rightarrow 1$, the perturbative
kernel determines the asymptotic $x$-dependence, while if the initial
conditions are singular, the asymptotic behavior is stable under evolution,
and in this sense is fundamentally nonperturbative (but see below).

In DIS, which is it?  
In fact, the question may not be so well posed, since
evolution is not fully determined by the $x\rightarrow 0$ behavior
of the gluon distribution.  In addition, the 
electromagnetic structure functions are not adequate
to determine the large number of parton distributions, including
valence and sea quarks.  Therefore, more than one parameterization
may, and does, fit the data.  
On the one hand, ``global" fitting groups, MRS \cite{mrs1,mrs2} and CTEQ, \cite{cteq}
with inputs from a wide range of  $x$ and $Q^2$ in a variety of
experiments, find that the increasing steepness of $F_2$
as $Q^2$ increases is correlated with a steepening gluon distribution,
 with an effective exponent $\lambda(Q^2)$ that obeys
$d\lambda/d\ln Q^2>0$.  This behavior is built into the
GRV parton distributions, \cite{grv} starting with a negative
value of $\lambda$ for small $Q^2$.  The actual evolution of 
each of these gluon distributions
is much more complex than simply case (i) or (ii).
On the other hand Lopez {\it et al.}\ have described a fit \cite{Yn} that
realizes case  (ii) in the sense that the leading power of $x$
in the gluon distribution is $Q^2$-independent.  In their
parameterization, $F_2$ softens at lower  $Q^2$ because of
a $Q^2$-dependent prefactor for $x^{-\lambda}$.  
In any case, the ability of
the factorization formalism to describe DIS down to
$x\sim 10^{-6}$ -- corresponding to a parton that
carries one one-millionth of the proton's momentum --
is a testimony to the extraordinary range
of the perturbative picture that generalizes the parton model.

In this context we may also observe that 
there is a perturbative evolution that
apprears to be kernel-driven.  
This is the MLLA (``modified
leading logarithm approximation") for partonic multiplicity
distributions in jets \cite{MLLA}.   Distributions for partons
are identified with those for hadrons; a method known as
``parton-hadron duality".  Such an identification is
particularly natural for a process in which evolution
is driven by a perturbative mechanism that may be  followed
from the shortest distances out to $1/\Lambda_{\rm QCD}$ 
self-consistently, although the nature of corrections to 
parton-hadron duality predictions are not fully understood.
MLLA evolution is quite different from DIS, however,
because in the former QCD interference
actually suppresses the emission of soft gluons.  
New data that compares well with the predictions of
the MLLA has been presented by OPAL \cite{Duchen} at this
conference.  

By the same token, we may ask whether it is possible to find
a singular initial condition like case (ii) above, which
is perturbative.  The answer is yes, at least at the
level of leading logarithmic approximations, in the
context of the BFKL treatment of DIS, to which we now
turn.

\subsection{BFKL Evolution}

The BFKL equation for DIS follows from a modified version of the factorized
expression, Eq.\ \ref{dglap}, in which the parton fractional momentum $\xi$
itself plays the role of a factorization scale.  To emphasize this point,
consider the following two factorized expressions for a  structure
function $F$ ($F_2$ for instance),
\bea
F(x,Q^2)&=& \int_x^1 {d\xi\over\xi}\; C\bigg({x\over\xi},{\mu^2/Q^2}\bigg)G(\xi,Q^2)\nonumber\\
&\ &\quad +{\cal O}\bigg ({1\over Q^2}\bigg )\nonumber \\
= &\int&d^2k_T\; c\bigg({x\over\xi'},Q,k_T\bigg)\psi(\xi',k_T)\nonumber\\
&\ &\quad +{\cal O}\left({1\over \ln (1/x)}\right )\, .
\label{bfklfact}
\eea
The first equality is the factorization Eq.\ \ref{factotimes} for 
$F$, specialized to gluons only, and represented by Fig.\ \ref{DIS}b above.
(In 
DIS, we ``cut" this diagram:  that is, separate it into amplitude and  complex
conjugate to form a cross section.)
The ``coefficient function" $C$ plays the role  of the
hard-scattering cross section $\hat{\sigma}_{\rm PT}$ in
Eq.\ \ref{factotimes}.  Corrections to this equality are
suppressed by $1/Q^2$ (neglecting for this argument the effects of quarks). 
We recall that in this expression we have organized momentum
scales larger that the factorization scale $\mu$ into $C$,
and those smaller into $G(\xi,\mu^2)$.

\begin{figure}
\centerline{\epsffile{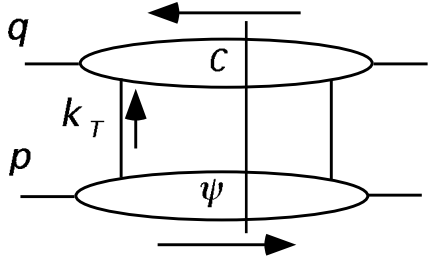}}
\caption{Factorization associated with BFKL resummation of logarithms
of $x$.}
\label{BFKLDIS}
\end{figure}

The second expression in Eq.\ \ref{bfklfact} is a reorganization
of the same structure function $F$, based upon a factorization
of partons with fractional momenta smaller than 
scale $\xi'$ into a new coefficient function $c$.
Partons of larger fractional momenta are absorbed into a new, long-distance
wave function $\psi$.  This factorization is illustrated schematically
in Fig.\ \ref{BFKLDIS}, in which the process consists of two ``jets" of
virtual particles, one set nearly parallel to $p$, the other
nearly parallel to $q$, which exchange a pair of quanta.  Note that the picture requires
that $Q^2=-q^2$ be much less than $p\cdot q\sim s$,  so that
$q^\mu$ may be thought of as having a definite direction. 
This is precisely the small-$x$ region in which we are interested.
The factorization is based on the observation that as $p\cdot q\rightarrow \infty$,
the assignment of quanta between the two ``jets" is somewhat arbitrary, and will
change as we sample frames that differ by boosts along the collision axis.

Now, instead of a convolution in 
fractional momenta, the convolution is in transverse momenta.
$\psi$ is related to the standard gluon distribution by
\be
G(x,\mu^2)=\int^\mu d^2k_T\; \psi(x,k_T)\, .
\ee
Despite the analogous structure of the two convolutions in
Eq.\ \ref{bfklfact}, corrections to the second form are
very different, suppressed only by inverse powers of $\ln(1/x)$.
Nevertheless, up to these corrections, both forms are simultaneously
valid.

Just as the first factorization form leads to DGLAP evolution by invoking the
independence of $F$ on $\mu$, so the independence of $F$ on $\xi'$
leads to another evolution equation, analogous to Eq.\ \ref{dglap},
but now with a convolution in transverse momentum rather than
parton fraction.  This is the BFKL equation, \cite{bfkl}
\be
\xi{d\psi(\xi,k_T)\over d\xi}=\int d^2k'_T\; {\cal K}(k_T-k'_T)\psi(\xi,k'_T)\, .
\nonumber\\
\label{bfkl}
\ee
The kernel $\cal K$ turns out to be independent of $\xi$ in this
approximation, although by dimensional analysis it could
have depended upon it.
Solutions to the BFKL equation may be found by substituting 
trial solutions in the form of powers of $\xi$ and $k_T$.
These solutions take a continuous range of powers of $\xi$,
of which the most singular is the famous BFKL result,
\be
\xi\psi(\xi,k_T)\sim\xi^{-4N\ln 2(\alpha_s/\pi)}\, .
\label{bfklpom}
\ee
This is precisely of the form of case (ii) above, and lends further
plausibility to that approach.

Concerning this result, however, we should make a number of
observations.  First, the BFKL equation for DIS is not in
contradiction to DGLAP evolution in general.  Indeed, as we
have seen above, it is simply based on a reformulation of the same
underlying factorization of long- and short-distances.  In this
sense, it picks out those terms that produce leading logarithms of $1/x$, to all orders
in the strong coupling $\alpha_s$, from the DGLAP kernel.  The question
arises, however, which $\alpha_s$?  
If we pick $\alpha_s(Q^2)$,  the phenomenology of small $x$
structure functions definitely disfavors Eq.\ \ref{bfklpom},
because this distribution would become less rather than more
steep as $Q^2$ increases.  
Because the BFKL equation
results from a purely leading log
analysis, however, the argument of the coupling
is not determined. Eventually,
the two-loop version of the kernel  $\cal K$ in Eq.\ \ref{bfkl} will
help, but there is another, even more important, consideration.
In the DGLAP equation, corrections
at second and higher order in $\alpha_s$ modify the kernels $P_{ij}$,
but do  not change the equation itself. Nonleading logarithms
in $1/x$, on the other hand, actually change the form of the BFKL equation, requiring
more $k_T$ integrals. \cite{bjkp} 
(At large $Q^2$, however, Eq.\ \ref{bfklfact} may be replaced
by a convolution in $k_T$ and $\xi'$. \cite{ktfact})
In the next section, we will
review some of the additional features of BFKL evolution that,
despite all this, make it such
an attractive subject of study.

\subsection{BFKL: the Larger Context}

DIS is only one application of the BFKL analysis.  
Others may be found by noting the relation of
the DIS distribution $\psi$ to
the total cross section for the
scattering of a nucleon by an off-shell photon.  In
the $x\rightarrow 0$ limit, $x$ is proportional to the ratio
of the squared invariant mass of the photon-nucluon system to
the ``mass" of the photon. Thus, starting with
\be
\psi\rightarrow s_{\gamma^*p}\sigma^{\rm tot}_{\gamma^*p}\sim s_{\gamma^*p}{1\over x^{1+\delta}}\, ,
\ee
where $\delta$ is given by the exponent in Eq.\ \ref{bfklpom},
we find
\be
s_{\gamma^*p} \sigma^{\rm tot}_{\gamma^*p}\sim\Bigg( {W^2\over Q^2}\Bigg)^{1+\delta}\sim 
\Bigg( {s_{\gamma^*p}\over m_{\gamma^*}^2}\Bigg )^{1+\delta}\, .
\ee
In this picture, the total $\gamma^* p$ 
cross section increases as a power of the relevant 
c.m.\ energy squared.  This result
is at the center of an entire literature.  In fact, it was derived first in
the context of hadron-hadron scattering, for which 
analogous reasoning gives \cite{bfkl}
\be
\sigma_{AB}^{\rm tot}\sim s^{\alpha-1}\, ,
\label{sigtotbeh}
\ee
with $\alpha>1$ for any two hadrons $A$ and $B$ to leading
logarithm in $s$.  

The result Eq.\ \ref{sigtotbeh} is interesting for (at least)
two reasons. First, because it is strongly reminiscent of
``Regge" behavior, which was long ago conjectured
for the $s\rightarrow \infty$, $t$ fixed limit in field theory,
\be
\sigma_{AB}^{\rm tot}(s,t) \sim \sum_i\beta_{Ai}(t)s^{\alpha_i(t)-1}\beta_{Bi}(t)\, .
\label{regge}
\ee
Here $i$ labels an exchanged ``reggeon", a composite degree of freedom.
The function $\alpha_i(t)$, when analytically continued to 
specific timelike values of $t$, 
is identified with the spin of the particle whose exchange describes the reggeon,
$\alpha_i(t=m^2_{iJ})=J$.  Although the pomeron's timelike continuation
is a matter of debate, we recognize the tantalizing possibility that here 
high-energy analysis has found a window to hadronic degrees of freedom.

The second source of interest in Eq.\ \ref{sigtotbeh} is a problem:
such a power behavior extended to arbitrarily large $s$ would eventually
violate unitarity bounds on the total cross section.  Thus, we are naturally
led to study corrections to Eq.\ \ref{sigtotbeh}, and, given the
accuracy of the factorization that leads to it, Eq.\ \ref{bfklfact}.

Current theoretical work on BFKL resummation and
the Regge limit in QCD may be classified roughly into two categories.
One might call the first ``visionary BFKL", which attempts to
grapple with the two fundamental
issues of Regge behavior and unitarity directly as mathematical
problems in the presence of strong
coupling, going far beyond perturbation
theory. This is to be contrasted with ``pragmatic BFKL", which 
attacks problems where the coupling is small, either
in model systems or in special kinematic limits.
 
Attempts to understand the BFKL equation 
in a nonperturbative fashion start with the observation that,
rewritten as
\be
{d\over d\ln x}\psi = {\cal K}\otimes \psi\, ,
\label{bfklotimes}
\ee
the BFKL equation (Eq.\ \ref{bfkl} above) is
similar to a Schr\"odinger equation, with $\ln x$
playing the role of time, and $\cal K$ of the Hamiltonian. \cite{lipatov1,fadkor}
For a solution with behavior $x^{-\alpha}$, 
the Regge exponent $\alpha$  (see Eq.\ \ref{regge})  plays
the role of an energy.  
As illustrated in Fig.\ \ref{BFKLDIS}, the wave function
is characterized by two external particles, which at lowest
order are gluons, while at higher orders have the interpretation
of ``reggeized" gluons -- corresponding to a gluonic
Regge trajectory in the general form of Eq.\ \ref{regge} \cite{regglue}.  
Evidently confinement prevents such an  exchange from contributing to physical
hadronic scattering, but its bound states can contribute, and the
simplest of these is the two-reggeon bound state, the pomeron,
which is the solution to eq.\ \ref{bfklotimes}.  From this
point of view, it is possible to generalize the Hamiltonian
$\cal K$ to ${\cal H}_n$, which describes the dynamics of $n>2$
reggeized gluons, at least when 
$N\rightarrow \infty$, with $N$ the number of colors.  
As discussed at this conference, \cite{lipatov2,korconf,wosiek} the
spectra of the ${\cal H}_n$ may be determined 
in principle by the
methods of statistical mechanics.  Quite recently, progress has been
made in determining the specific ``energy" of the three-reggeon
bound state, known as the ``odderon".  \cite{korconf}
If, as suggested at this conference, the odderon has an exponent
above unity, then it might be possible to detect it experimentally. \cite{predazzi}
In the long run, if all goes well, and all
the $\alpha$'s are found as functions of $t$, one may analytically
continue and really solve for the bound states of QCD.  This
vision is far from realization, however, and we should emphasize
that the construction of the multireggeon Hamiltonians requires
approximations, such as large numbers of colors,  whose
validity it is currently difficult to judge.

\begin{figure}
\centerline{\epsffile{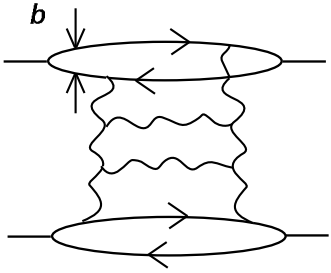}}
\caption{Gluon ladder exchange in onium-onium scattering.}
\label{onia}
\end{figure}

On the pragmatic side, a laboratory for studying
the BFKL pomeron in a perturbative context is in the
hypothetical scattering of heavy-quark ``onia", as illustrated
in Fig.\ \ref{onia}. \cite{onia} Because the size $b$ of a $Q{\bar Q}$ bound state
shrinks with the quark's mass, such a bound state behaves
as a ``color dipole", and $b$ as an infrared cutoff
on the running of the coupling.   In this case, the sum of
ladders, as in the figure, may be studied perturbatively, because we
may arrange for $\alpha_s(1/b)$ to be small,
while $\alpha_s\ln s$ is large.  In addition, in the large-$N$
limit, the gluons themselves act as color dipoles, and the 
process of ladder formation becomes iterative, allowing for an
explicit construction of the relevant wave functions.  This
model serves as an alternate approach to generalize the 
leading-logarithm, BFKL equation, and to study how unitarity
is realized.  

There is, in addition, a series of proposals for observing
the BFKL pomeron phenomenologically, in hard, high-energy
scattering.  In each case, a process has been identified in
which logarithms of $s$ (analogous to $1/x$, as we have observed
above) are generated, with an infrared cutoff set 
in the perturbative region by a momentum transfer
$Q$, with $\Lambda_{\rm QCD}\ll Q \ll \sqrt{s}$.  Examples
are in the rapidity-dependence and angular correlations of
jets of transverse momentum $Q$ in very high-energy 
hadronic collisions. \cite{dds,fehr,munav} The masses of virtual photons in 
$\gamma^*\gamma^*$
scattering at very high energy
also make it possible to derive an infrared safe, BFKL-resummable 
cross section. \cite{bhs} Generally speaking, however, this sort of experimental signal for BFKL
has not turned out to be easily accessible, in large
part because the leading-logarithm approximation requires a
very high energy to provide numerically reliable predictions.

Despite the limitations of the quantitative predictions of 
BFKL resummation, there remain a number of experimental
hints that the physics behind it is relevant.  At this conference,
we heard, for instance, about the distribution of particles emitted
as a function of transverse momentum for various ranges of $x$. \cite{mroczko}
In this connection, we may also mention treatments of DIS final
states based on coherence analysis and ``angular ordering",
\cite{CCFM} which allow for interpolations at the leading-log
level between BFKL and DGLAP equations.  
The lack of transverse-momentum ordering in BFKL evolution (see below)
suggests that it will produce a ``harder tail" at high $p_T$ 
than might be the case for (leading or next-to-leading logarithm) DGLAP evolution. \cite{kms}
H1 reported some indications of this effect in their 1994 data, although
the interpretation is based on 
comparison to  the outputs of event generators.
The other experimental signals that we may connect to
the program of color singlet exchange are the 
``rapidity gap" events, which comprise measureable fractions
of jet cross-sections at both HERA and the Tevatron. \cite{devlin,zylberstejn}
In a rapidity gap event, a pair of jets is produced at very high
$p_T$, with very little radiation between them, a phenomenon
that is most naturally attributed  to the exchange of 
some sort of color singlet. \cite{csing}
A full perturbative treatment has not, so  far as I  know,
yet been given for these processes, and remains
as a challenge for the coming years.
As the overall momentum transfer vanishes compared to 
$s$, rapidity events merge smoothly into ``diffractive" events,
on which there is now much more data from DIS,  and 
toward which much more theoretical speculation has been directed. 
It is to this constellation of signals that we now turn.

\section{Evolution and Hard Diffraction}

\subsection{Interpreting Evolution}

One of the most attractive features of DIS is found
in the interpretation of its final states as
virtual states of the proton ``brought to life" by
the energy provided by the lepton.  By inspecting distributions
of hadrons in
the final state, we  may hope to learn, not only about
the distribution of scattered partons, but also about
the sequence of virtual processes by which they are produced.
Such a linkage of hadrons to partons takes us beyond the
strict application of ``infrared safety", in which quantities
are computed in perturbation theory and applied directly to
inclusive cross sections.  Rather, we shall apply the still not
fully quantified concept of parton-hadron duality,
referred to above.

To discuss distributions of final-state partons, we note
that evolution equations arrive at (perturbative) final states
by creating many short-lived  states from fewer long-lived
states.  A highly-virtual exchange, such as in DIS, 
produces a cross section that is dominated by short-lived
states,  simply because there
are so  many of them.  Nevertheless, the dominant contribution
to a short-lived state contains many footprints of the
path taken to arrive at that state.  The DGLAP and BFKL equations, the primary
contenders for describing DIS at low $x$, each has a
characteristic ``history", embedded within its final states.
Let us be clear, however, that in this context 
by ``DGLAP evolution", we mean evolution according to 
Eq.\ \ref{dglap} with a finite-order approximation to the
DGLAP kernels $P_{ij}$.  BFKL evolution
is consistent with all-orders DGLAP evolution, but has
certain qualitative differences from its low-order approximations.

For DGLAP evolution defined in this sense, 
final states are characterized by partons with ordered $k_T$,
with a moderate decrease in $x$ as $k_T$ increases.  Hadrons with
the largest $k_T$ are descendants of those partons emitted
at the last stage of evolution.  BFKL evolution, on the other hand,
is characterized by a rapid decrease in $x$, coupled with slow,
perhaps random, variations in $k_T$.
The distinction between these two pictures has an intruiging
heuristic
relation to the diffractive events, observed with 
surprising frequency at HERA, and at fixed-target
experiments. \cite{levonian}

\subsection{HERA ``Stories"}

\begin{figure}
\centerline{\epsffile{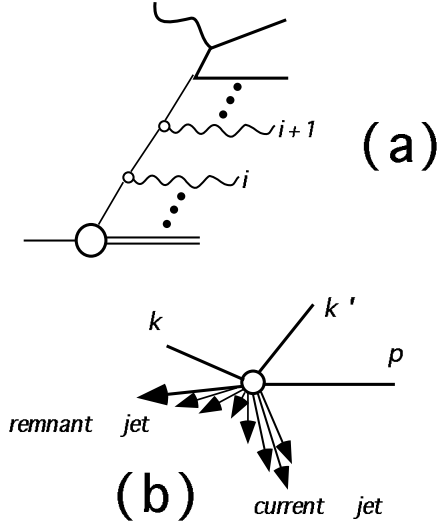}}
\caption{DGLAP story of evolution.}
\label{DStory}
\end{figure}

The relation of evolution to distributions in the final state
may be illustrated heuristically by two ``stories" of evolution, corresponding
to  DGLAP and BFKL evolution.  The first, a  DGLAP-like history, 
is illustrated by Fig.\ \ref{DStory}a, in  which partons
are emitted  in a ladder strongly-ordered in
transverse momenta, $k_{i+1,T}\gg k_{i,T}$.
Assuming parton-hadron duality, and ignoring perturbative
final-state interactions, suppressed by factors of at 
least $\alpha_s(k_{i,T})$, we anticipate a distribution of
final-state hadrons that reflects the distribution of the
partons that emerge from the ladder.  Because of the $k_T$-ordering,
the hadrons spread out in the final state, filling a range of directions
between the scattered quark (the ``current jet") and the unscattered
partons (the ``proton remnant").  
In the c.m.\ frame such a final state would appear
as in Fig.\ \ref{DStory}b.  In particular, 
we expect such a process to 
have no large angular or rapidity gaps in the final state, because
there are no large gaps in a leading-log ladder, and because
the partons are rapidly separating from each other  as  they
propagate into the final state.

\begin{figure}
\centerline{\epsffile{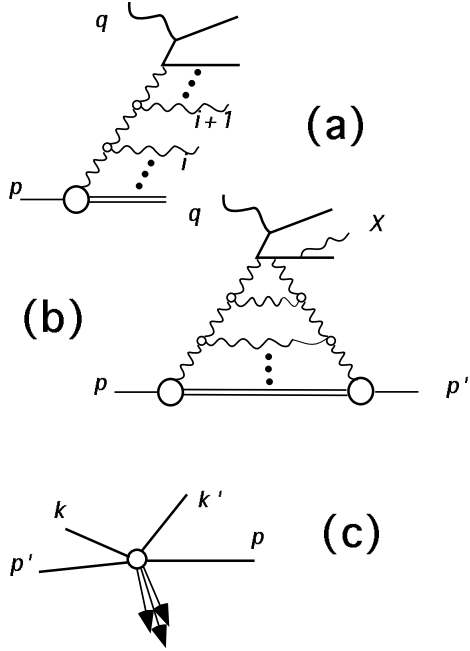}}
\caption{BFKL story of evolution}
\label{BStory}
\end{figure}

To the above picture we may contrast the ``BFKL-like" story of
Fig.\ \ref{BStory}a-c. Here, (5a) the ladder consists of
partons emitted with comparable transverse momenta, $k_{i,T}\sim k_{i+1,T}$,
but highly-ordered longitudinal momenta, $x_i\gg x_{i+1}$.  If, in
addition, none of these (randomly distributed) transverse momenta 
is too large, then there may well be a substantial spatial overlap of
the radiated ladder gluons with each other {\it and with the
target proton}.   Finally, if the ladder gluons are  in a singlet
state, \cite{bucheb} which costs them only the emission of one extra gluon, they
may have a substantial overlap with the proton remnant to form
an outgoing proton, or nucleon resonance (\ref{BStory}b).
Since the  ladder gluons, except perhaps for some at the
very top of the ladder, do not appear in the final state,
this corresponds to an event of the type  shown in \ref{BStory}c,
in which the current jet is isolated, with a  large ``gap"
in rapidity between it and the scattered proton $p'$.

The ladder of virtual gluons plays the
role of a ``pomeron".  The pomeron in this picture of
diffractive scattering is a process, rather than a particle.
Nevertheless, it is natural to interpret diffractive hard
scattering as being initiated by a parton, not simply
from the proton, but from the ladder, or pomeron.  In this
interpretation, we reduce the diffractive scattering into
a two-stage process: the (effective) emission of a pomeron
from the proton,
with fraction $x_{\cal P}$ of the proton's momentum, followed by
a hard scattering of a parton with a fraction $\beta$ of
the pomeron's, and hence fraction $\beta x_{\cal P}$ of the
proton's, momentum. \cite{ispict} The variable
$\beta$ may be determined from the invariant mass 
$M_X$ of the current jet, through the standard relation,
\be
M_X^2={Q^2\over\beta}(1-\beta)=(x_{\cal P}p+q)^2\, ,
\end{equation}
or, equivalently, $\beta=x/x_{\cal P}$, with $x$ the usual
scaling variable.

If this partonic interpetation is to be valid, 
it is also plausible to explore the possibility of 
approximate scaling for hard diffractive scattering, with
calculable evolution. This expectation has 
been formalized \cite{bersop} in a generalization of DIS factorization
to inclusive structure functions for the diffractive
cross section (compare Eq.\ \ref{factotimes}),
\bea
F_2^D&=&C_{2,i}^D\otimes\phi_{i/p}^D\nonumber \\
&=&\int_\beta^1{d\beta'\over\beta'}C_{2,i}^D(\beta/\beta')
\phi_{i/p}^D(\beta')\, .
\label{ftwod}
\eea
If this factorization holds, $F_2^D$ evolves according to DGLAP
evolution, but with yet another initial condition.
The H1 collaboration,
by performing an analysis of $Q^2$ dependence for
different values of $\beta$, has concluded
 that the distribution of partons
in the pomeron may be concentrated near unity in $\beta$, 
in distinction to normal hadronic parton distributions. \cite{phillips}

At the same time, 
because this effective pomeron appears in both the amplitude
for this process and its complex conjugate,
the dependence of the diffractive 
structure function on the {\it total} hadronic
energy $W=\sqrt{(p+q)^2}$ behaves as the {\it square}
of a total cross section,
\be
F_2^D(W)\sim ({\rm kinematics})\times 
\bigg(W^2\bigg)^{2(\alpha_{\cal P}-1)}\, .
\ee
Here $\alpha_{\cal P}$ is the power of $s$ characteristic of
single-``pomeron" exchange.  If this pomeron were the same
at all momentum transfers, it would be directly
 related by the
optical theorem to the exponent in
Eq.\ \ref{sigtotbeh} for the total hadronic cross section.  
Experimentally, however, the exponent $\alpha_{\cal P}$ 
depends upon the momentum transfer, and $F_2^D$ increases
faster with $W$ when $Q^2\gg\Lambda_{\rm QCD}^2$ than in the
photoproduction limit. \cite{phillips,barbagli}
Thus, it may well be useful to distinguish between
two ``pomerons", a ``soft" pomeron that generates the total
cross section Eq.\ \ref{regge}, with a smaller exponent $\alpha_{\cal P}\sim 1$, 
and a ``hard" pomeron that
couples to high-energy scattering processes, with a somewhat
larger exponent, perhaps suggestive of the BFKL result, Eq.\ \ref{bfklpom}.  \cite{landshoff}
It is not entirely clear, however, whether 
the experimental results are consistent with a purely pomeron-like
picture. \cite{phillips}

\subsection{Exclusive Electroproduction}

\begin{figure}
\centerline{\epsffile{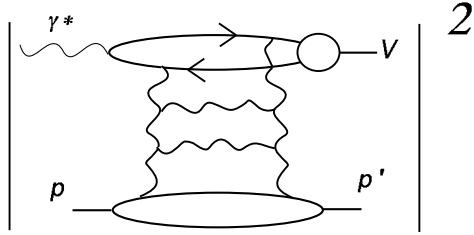}}
\caption{Exclusive electroproduction.}
\label{EE}
\end{figure}

Further insight into diffractive processes may be gathered
by demanding an {\it exclusive} final state, as in
electroproduction of vector bosons, 
$\sigma_{{\gamma p}\rightarrow Vp+X}$, with $V=\rho,\; \psi$, 
etc., at photon virtuality $Q^2\gg \Lambda_{\rm QCD}^2$.
  The price of demanding an exclusive final state is that
the cross section is power-suppressed in $Q^2$.  The 
leading-logarithm process is illustrated in Fig.\ \ref{EE},
in which the gluonic ladder structure is manifest at
the amplitude level.  Identifying the ladder via unitarity
with the gluon distribution, again valid to leading logarithm,
one finds, \cite{exelsigma,teubetal}
\be
{d\sigma\over dt}\sim {\alpha_s^2\over Q^6}\big |xG(x,Q^2)\big|^2\, ,
\label{vectboson}
\ee
where the constant of proportionality depends on the 
quark-antiquark wave function of the vector boson.
In Eq.\ \ref{vectboson}, $x\sim 1/W^2$ by the usual kinematics.
If $xG(x)\sim x^{-\lambda}$,
then $d\sigma/dt\sim W^{4\lambda}$, and this cross section,
although suppressed in $Q$, increases dramatically with
$W$.  Such an increase has, indeed, been observed. \cite{exelexp}
At the same time, we should note that the gluon ladder
in Fig.\ \ref{EE} is evaluated at $t=(p-p')^2<0$,
while the true gluon distribution requires $t=0$
exactly.  It is possible, however, to analyze this
approximation systematically. \cite{rady}

\section{High-$p_T$, Heavy Quarks and Resummation}

\subsection{Jet and Heavy Quark Cross Sections}

Jet physics is coming of age.
Presentations at the parallel sessions on hard processes and perturbative QCD
described an increasingly sophisticated analysis of jet 
cross sections.  At HERA, ZEUS and H1, at LEP, the ALEPH, DELPHI OPAL and L3 collaborations,
and at SLC, the SLD collaboration,
have measured multijet 
cross sections, analyzed jet-related event shapes and fragmentation functions.  The LEP collaborations reported on
methods for routinely distinguishing quark and gluon jets.  CDF and D0 are also
having a close look at jet substructure and multijet events.  I have space here for only
a few of the many physics issues raised by these important investigations.

High-$p_T$ jet cross sections at hadron
colliders are one of the proverbial
successes of perturbative QCD. \cite{nlojets,sopertalk}  Within
the foregoing year, however, jet cross sections at the highest
energies received a new burst of attention with the 
description by CDF \cite{devlin,cdfjets} of a substantial, and apparently growing, excess
of events at the highest transverse momenta, up to approximately
one half of a TeV. D0 does not see such a rise, \cite{dojets} although the uncertainties
in both experiments are such that they are not in strong disagreement.
 This conference has seen a general consensus
that this is an inconclusive signal of new physics, and may be more a
reflection of remaining uncertainties in parton distributions, \cite{brocktalk} particularly
the gluon distribution \cite{cteqjetpdf,mrs2} at relatively large 
values of fractional momentum. \cite{sopertalk}
With suitably-adjusted parton distributions, the now-classic next-to-leading
order (NLO) formalism may well provide a good picture of  the data, at least
at the current level of statistical and systematic errors.  

Another, and closely related, subject of keen interest over the
past year has been the computation of the total cross section for
the production of top quarks in hadronic collisions, particularly
at the Tevatron. \cite{topqkexp} Here, the most widely-quoted analyses go beyond
NLO, resumming ``large", positive corrections at ``partonic threshold"
to all orders in perturbation theory.  \cite{lsvn,bc,cmnt} A number of viewpoints
have recently been raised on the best way of going about this resummation,
and whether or not it results in substantial modifications of
NLO results.  

The outcome of these discussions is relevant to
our confidence in the NLO cross
sections for jet and heavy particle production, and hence to
signs for new physics.  Indeed, as we shall see,
the perturbative calculation of hard-scattering cross sections
through factorization leads to a
nonconvergent series in the coupling.
This problem has a  resolution, however, that is surprisingly consistent with the
successes of NLO approximations, and which contains within it a connection to an as-yet 
undeveloped formalism for power corrections in general QCD
cross sections.  To try and clarify these issues,
I shall first discuss the origin  and resummation of ``Sudakov"
corrections in factorized cross sections.

The cross sections for jet production, inclusive top production,
and any other final states
$F$ of mass $Q$ in hadron (A)-hadron (B)
scattering take the detailed form, \cite{cssrv}
\bea
{d\sigma_{AB\rightarrow FX}\over dQ^2dy}
&=&
\sum_{ab}
\int_{Q^2/S}^1dz\; 
 \int\frac{dx_a}{x_a}\frac{dx_b}{x_b}\nonumber \\
&\ &\times \phi_{a/A}(x_a,Q^2) 
\phi_{b/B}(x_b,Q^2) \nonumber \\
&\ & \times \delta\left(z-{Q^2 \over x_ax_bS}\right)\; 
\nonumber \\
&\ &\times {\hat \sigma}_{ab\rightarrow FX}\left (z,y,x_a/x_b,
\alpha_s(Q^2)\right)\, , \nonumber \\
\label{basicfact}
\eea
where the perturbative short-distance function begins with the
Born cross section.
\be
\hat{\sigma}=\sigma_{\rm Born}+{\alpha_s\over\pi}\hat{\sigma}^{(1)}
+\dots\, .
\ee
In the following, we shall discuss briefly the nature of 
corrections in $\hat\sigma$ beyond one loop, and how they 
have been used to 
improve upon NLO predictions.  We shall see once again
that factorization plays a central role in this resummation.

\subsection{Partonic ``Threshold" and Noncovergence }

Partonic threshold in Eq.\ \ref{basicfact} refers to the limit $z\rightarrow 1$,
in which the total partonic c.m.\ energy $\sqrt{x_ax_bS}$ is just large enough
to produce the observed final-state system $F$, with no energy left over
for QCD radiation.  (In the production of heavy 
quarks, partonic threshold should be distinguished
from ``true" threshold in which the pair is produced at rest, $Q=2m_q$.) 
Not surprisingly,  partonic threshold is a singular limit, since,
even classically, hard
scattering demands copious radiation.

The singular nature of partonic threshold is 
reflected in the presence of potentially large corrections.  At $n$ loops,
the dominant ``large" correction is the ``plus distribution",
$-{\alpha_s^n\over n!}\times$ $\times\left [ {\ln^{2n+1}\left((1-z)^{-1}\right)\over 1-z} \right ]_+$,
which is defined by its integrals with smooth functions ${\cal F}(z)$ (such as the parton
distributions in Eq.\ \ref{basicfact}), through,
\bea
&-&{\alpha_s^n\over n!}\int_0^1dz\; 
{{\cal F}(z)-{\cal F}(1)\over 1-z}\; \ln^{2n+1}\left((1-z)^{-1}\right) \nonumber \\
&=&{\alpha_s^n\over n!}\int_0^1dz\; {\cal F}'(1) \ln^{2n+1}\left((1-z)^{-1}\right)+\dots \nonumber \\
&\sim& {\cal F}'(1) {\alpha_s^n\over n!}{(2n+1)!}+\dots\, ,
\label{factorial}
\eea
where ${\cal F}'(1)$ is the first derivative to ${\cal F}$ at $z=1$.  
At $n$th order, such a correction produces a
coefficient of $\alpha_s^n$ that grows faster than  $n!$. These corrections
have the same sign for every value of $n$, and for $n$ high enough are large,
positive, and  nonconvergent.  

\begin{figure}[ht]
\centerline{\epsffile{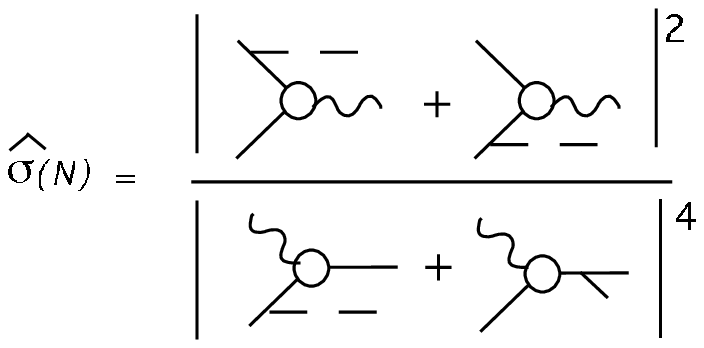}}
\caption{Schematic representation of moments of the
Drell-Yan partonic hard-scattering function.}
\label{mtratio}
\end{figure}

The origin of such terms is in the process of
factorization itself, as it is normally implemented.  The issue may be
illustrated schematically as in Fig.\ \ref{mtratio}
for $\hat{\sigma}$ in the Drell-Yan (DY) cross section.   The hard-scattering
function $\hat{\sigma}$ is computed  from parton-parton cross sections by
factorizing, that is, dividing out, distributions for each 
incoming parton.  But these distributions, the denominator of the figure,
are {\it each}
the square of an amplitude, in which there are both incoming {\it and}
outgoing partons, both of which have a tendency to radiate.   In 
the numerator, however, we need only worry about the radiation
of the incoming partons.  
This difference is also illustrated by comparing the invariant masses of the hadronic
final states in the DY and  DIS processes,
\bea
{\rm DY}&:&\quad W^2_{\rm had}=Q^2(1-z)^2\nonumber \\
{\rm DIS}&:&\quad W^2_{\rm had}={Q^2(1-x)\over x}\, .
\label{txtoone}
\eea
In the former, there is a quark-antiquark pair in the initial state
only, while in the latter, which defines a
distribution in ``DIS scheme", a quark or  antiquark is present with
large energy in the final state as well.  As we shall see, both the numerator 
and denominator -- the DY and DIS cross sections -- vanish at threshold.
But the denominator, with {\it two} factors of the DIS cross section,
vanishes faster, and this results in an enhancement in $\hat{\sigma}$.
To resolve this problem in a general case, we need to develop
a method for organizing large corrections in the elastic limit.
This is known as Sudakov resummation, and like DGLAP evolution 
and BFKL resummation, it is based on a factorization property
of QCD cross sections. \cite{cls}

\subsection{Sudakov Resummation}

\begin{figure}[ht]
\centerline{\epsffile{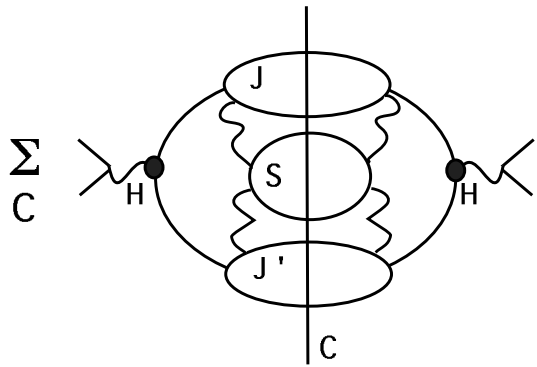}}
\caption{${\rm e}^+{\rm e}^-$ annihilation in the $T\rightarrow 1$
limit.  All lines in the $J's$ are mututally collinear, those in
$S$ carry vanishing momenta.}
\label{Ttoone}
\end{figure}

Sudakov resummation applies to a large class of hard-scattering
cross sections in ``exclusive" limits, 
for instance, $w\rightarrow 0$, when \cite{Tresum,disdyresum,qtresum}
\bea
w &=& 1-T\quad  {\rm e}^+{\rm e}^- \nonumber\\
  &=& 1-x \quad {\rm DIS} \nonumber \\
 &=& 1-Q^2/s\quad {\rm DY\ total}\nonumber \\
   &=&   Q_T/\sqrt{s} \quad {\rm DY\ or\ 2\ Jet}\, ,
\label{list}
\eea
where $T$ is the thrust variable.  
In the limit that any of these variables vanishes, all the 
final-state and/or initial-state particles in these processes
are either part of one or two massless {\it jets}, or have
vanishingly soft momenta.  In terms of momentum flow, these configurations are
identical to exclusive processes.
For the case of thrust, the cross section is illustrated in Fig.\ \ref{Ttoone}
in cut diagram notation, as a
sum over possible final states $C$ of an amplitude (to the left
of $C$) times its complex conjugate (to the right).

In this case, $w=1-T$ is the kinematic variable that vanishes 
in the two-jet limit for total c.m.\ energy $Q$.  
The singular part of $d\sigma/dw$ as $w\rightarrow 0$ 
enjoys a factorized form that
generalizes Eq.\ \ref{factotimes}, and in which 
the cross section is a convolution of functions,
one for each jet $J$ and $J'$, and one  for the soft radiation $S$,
linked by their contributions to the weight $w$, \cite{cls}
\bea
&&{d\sigma \over dw}
=
H(p_i/\mu,\zeta_i,\alpha_s(\mu^2)) \nonumber \\
&&\quad\times \int {dw_1\over w_1} {dw_2\over w_2} {dw_s\over w_s} \delta(w-w_1-w_2-w_s)\nonumber\\
&&\quad\times S(w_sQ/\mu,v_i,\zeta_i) \nonumber \\
&&\quad\times J(p_1\cdot \zeta_1/,w_1Q^2,\mu) 
J'(p_2\cdot \zeta_2,w_2Q^2,\mu)\nonumber \\
&&\quad\quad+{\cal O}(w^0)\, .
\label{sudfact}
\eea
Multiplying the convolution is an overall factor $H$, which summarizes
the effects of quanta that are off-shell by order $Q^2$.  Here, in addition
to a factorization scale $\mu$ that separates 
hard and soft quanta, we must introduce two new vectors
$\zeta_1$ and $\zeta_2$, to distinguish quanta moving in the
jet directions.   These may be thought of as gauge-fixing vectors.
Finally, the $v_i^\mu$ are lightlike vectors in the jet directions.
Once again, our freedom in choosing $\mu$, $\zeta_1$ and $\zeta_2$ 
enables us to derive a  useful resummation.  

As is often the case, resummation is most easily carried out in
a space of moments, under which the convolution, Eq.\ \ref{sudfact},
reduces to a product,
\bea
&&\int dw\, e^{-Nw}\, 
{d\sigma \over dw}
 \nonumber\\
&& \quad =
H(p_i/\mu,\zeta_i,\alpha_s(\mu^2))
{\tilde S}(Q/N\mu,v_i,\zeta_i)
\nonumber \\
&&\quad\, \times
{\tilde J}_1(p_1\cdot \zeta_1,Q^2/N,\mu) 
{\tilde J}_2(p_2\cdot \zeta_2,Q^2/N,\mu)\, .
\nonumber \\
\label{moments}
\eea
The independence of $d\sigma/dw$ from $\zeta_1$, for instance,
requires that 
\bea
&&p_1\cdot \zeta_1{\partial \over \partial p_1\cdot \zeta_1}
 \ln {\tilde J}_1(p_1\cdot \zeta_1,Q^2/N,\mu) 
\nonumber \\
&& \quad =
-p_1\cdot \zeta_1{\partial \over \partial p_1\cdot \zeta_1}
 H(p_i/\mu,\zeta_i,\alpha_s(\mu^2)) \nonumber \\
&& \quad \quad   -p_1\cdot \zeta_1{\partial \over \partial p_1\cdot \zeta_1}
 {\tilde S}(Q/N\mu,v_i,\zeta_i)\, ,
\eea
which is readily turned into an evolution equation for $N$-dependence.  
The process
is closely analogous to the procedure that leads from the $\mu$-independence of
Eq.\ \ref{factotimes} to the DGLAP evolution equations, \ref{dglap}.

The result of this procedure is an exponential of a double integral over
the running coupling at variable momentum scale, \cite{cls}
\bea
&&\ln \Bigg ({d{\tilde \sigma}(N) \over dw}\Bigg )
=
D_\sigma(\alpha_s(Q),\alpha_s(Q/N))
\nonumber \\
&& \quad  +
 2\int_{Q/N}^{Q}{d\lambda \over \lambda}
\bigg [
\int_{Q^2/\lambda N}^\lambda {d\xi \over \xi}\; A_\sigma(\alpha_s(\xi^2))
\nonumber \\
&& \quad\quad 
+ B_\sigma(\alpha_s(\lambda)) \bigg ]\, ,
\label{sudexp}
\eea
with $D$, $A$ and $B$ power series with infrared finite coefficients,
which can be determined at low order by direct comparison with
explicit calculations.  When the couplings are treated as fixed,
Eq.\ \ref{sudexp} produces an exponential of the form,
$\exp[-{\rm const}\times\alpha_s\ln^2 N]$.  The  running of the 
coupling produces ``subleading" logarithms, 
$\alpha_s^n \ln^mN$, with $m\le n$, in the exponent.
This exponentiated moment must be inverted to apply the
result to Eq.\ \ref{basicfact} in a practical case.  
At very large $N\ge Q/\Lambda_{\rm QCD}$, however, Eq.\ \ref{sudexp} becomes ill-defined,
as the perturbative coupling $\alpha_s(\xi^2)$ diverges.  We shall
return to this problem below.  Analogous reasoning applies to 
both the DIS and DY cross sections in their exclusive limits,
and leads to the Sudakov {\it enhancement} in the factorized
hard-scattering function at partonic
threshold noted above.

\begin{figure}[ht]
\centerline{\epsffile{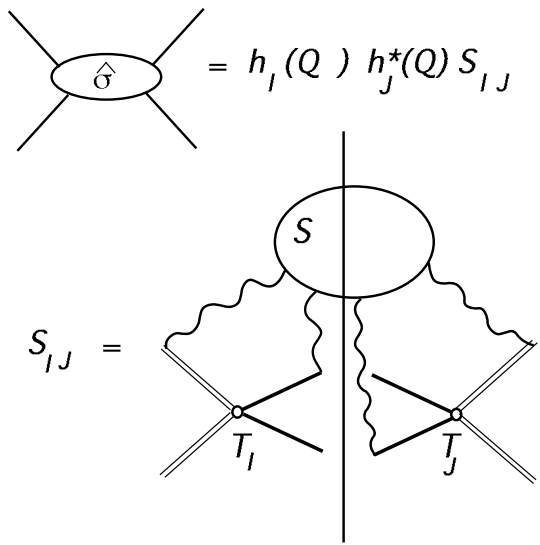}}
\caption{Representation of the factorization
of the hard scattering function $\hat{\sigma}$ 
near the elastic limit.}
\label{HS}
\end{figure}

The simple exponentiation of Eq.\ \ref{sudexp} applies,
unfortunately, only to QCD corrections to electroweak hard-scattering.  
For  QCD hard-scattering, such as jet and heavy-quark
production, which involve color
exchange, the exponentiation becomes a matrix problem,
which, however, is interesting in its own right. \cite{Lipatmx,BSSS,KK}
The methods required to treat this problem
turn out to be related, in particular, to the popular
``effective theory" approach to heavy quark physics.
Consider, for example, heavy quark production.
In this case, the dynamics of soft gluons ($S$ in Fig.\ \ref{Ttoone}),
may be represented as in Fig.\ \ref{HS}, in
which the hard-scattering function $\hat{\sigma}$ from
Eq.\ \ref{basicfact} is factored into two short-distance
functions $h_I$ and $h_J$, labelled by the
color exchange that takes place at short distances,
and a soft function $S_{IJ}$, which is the cross section for
eikonalized light quarks to annihilate to form 
eikonalized heavy quarks. \cite{KiSt} Here ``eikonalized" means
that these quarks emit gluons without recoil, and that
correspondingly their propagators are of the form
$1/v\cdot k$, with $v^\mu$ the velocity of the quark,
familiar from heavy quark theory.
The relevant evolution equations may be interpreted
as renormalization group equations for the composite
vertices that link the eikonalized quarks.  At {\it next-to}-leading
logarithm, the cross section in moment space is a sum over
color exchanges, of different exponentials, \cite{KiSt}
\bea
\hat{\sigma}_{q{\bar q}\rightarrow Q{\bar Q}}(N)
&=&\sum_{IJ}S_{IJ}\; h_I(Q)h^*_J(Q)\nonumber \\
&\ &\times\; e^{E_{IJ}(N,\alpha_s)}\, .
\label{qqQQexp}
\eea
At leading logarithm, $E_{IJ}$ 
is the same as in Eq.\ \ref{sudexp},
but beyond leading logarithm depends on
the color exchanges in the hard scattering.
Similar composite operators 
characterizing color exchange also occur in treatments
of hadron-hadron elastic scattering \cite{BSSS,KK} and the 
the decay and production of quarkonia. \cite{BBL}

Up until now, top production is the only 
inclusive cross section to have been studied extensively with
threshold resummation taken into account, and then
only at the level of leading logarithmic accuracy.
\cite{lsvn,bc,cmnt} (Very recently, however, 
a preliminary investigation of nonleading
logarithms has appeared. \cite{KiSV})
The various groups agree on the basic exponentiation
in $N$ space; differences in their results primarily concern
how best to invert the transform.  Especially
interesting is the question of what to do with
the running coupling in Eq.\ \ref{sudexp} when
$\xi^2_{\rm min}=Q^2/N^2\rightarrow \Lambda_{\rm QCD}^2$.
We shall touch on this issue in the  next section.

At the current 
experimental accuracy, there is actually little to
choose between the different numerical predictions.
In future Tevatron runs, however, we may hope for
more accurate measurements of the total top and 
jet cross sections, and by then some of the 
issues of numerical accuracy should be resolved.
In the meantime, it is worth noting that 
the only available (and
conservative) estimate of jet cross sections \cite{cmnt} shows the
resummed jet cross section rising, relative to NLO,
at the very highest $p_T$'s yet measured, by the order
of five percent.  The effect may be too small for 
current statistical errors, but it is an intriguing
contribution, which
requires an understanding of the theory
beyond the lowest orders.   Before passing to the
issue of how to deal with the running coupling, we
should note that a fullly consistent use of resummed hard scattering
functions requires a set of parton
distributions derived from experiment by employing the
same resummed hard parts.  It may take a little while
for someone to summon the courage to undertake such
a project.

\section{Power Corrections and IR Renormalons}

Even at the highest energies, precision 
for unaided NLO calculations is still relatively rare,
and this, as in the examples above, has stimulated the
development of Sudakov resummation techniques.  
We have already observed, however, that in Sudakov
resummation, going to all orders opens the
door to singularities associated with the divergence
in the running coupling, as in Eq.\ \ref{sudexp} when
the moment variable $N$ reaches $Q/\Lambda_{\rm QCD}$.  
We may  rely on event generators to build in
a model of high order corrections, but 
it's hard to know to what the nonperturbative inputs
in these models correspond in the true theory.
Indeed, the inclusion of any fixed dimensional
scale $\Lambda'$ in an event generator leads
to power-suppressed corrections of the form
$(\Lambda'/Q)^a$, $a>0$, in hard-scattering cross sections at 
scale $Q$. \cite{webber} All this has led a group
of investigators to ask whether there might
be a theory of power corrections in QCD cross sections.
\cite{shapeiir,dyirr,korst}  These 
and related ideas were discussed extensively
at the conference. \cite{irrpre,beneke}

Of course, there are famous cases in which a  
theory of power corrections already exists, based on the operator product
expansion (OPE).  For instance, the
total ${\rm e}^+{\rm e}^-$ annihilation cross
section is given by
\be
\sigma_{\rm tot}^{{\rm e}^+{\rm e}^-}
={(4\pi \alpha)^2\over Q^2}{\rm Im}\; \pi(Q^2)
\ee
where the function $\pi$ is found from the products of electromagnetic
currents,
\be
\pi(Q^2)=\left({-i\over 3Q^2}\right)\; \int d^4x\; {\rm e}^{-iq\cdot x}\langle 0|T j^\mu(0)j_\mu(x)|0\rangle\, ,
\ee
which at short distances (large $q^2$) may be expanded in terms of
local operators,
\bea
&&\langle 0|j^\mu(0)j_\mu(x)|0\rangle 
= {1\over x^6}C_0(x^2\mu^2) \nonumber \\
&& \quad +{1\over x^2}C_{F^2}(x^2\mu^2) \nonumber \\
&&\quad \quad \times \alpha_s(\mu^2)\langle 0|F_{\mu\nu}F^{\mu\nu}(0)|0\rangle
+\dots\, .
\label{vevope}
\eea
Here we have shown only two terms, the leading power, 
found from perturbation theory with vanishing quark masses ($\sigma_{\rm tot}$, of
course, is infrared safe), and the ``gluon condensate" term, suppressed
by $(x^2)^2$ compared to perturbation theory at short distances, but still singular.
Suppose we had never heard of the OPE; could we have discovered the $F^2$
term in Eq.\ \ref{vevope} on the basis of perturbative arguments alone?  

\begin{figure}[ht]
\centerline{\epsffile{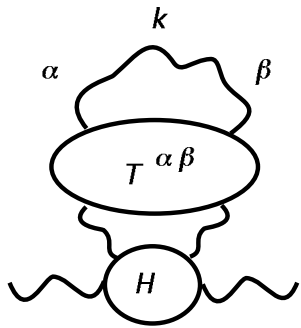}}
\caption{Subintegral that gives rise to nonconvergent
behavior in $\sigma_{\rm tot}^{{\rm e}^+{\rm e}^-}$.}
\label{irree}
\end{figure}

In fact, we could have done so, by looking at the large-order
behavior of coefficients $C_0^{(n)}$ of $\alpha_s(Q)^n$. \cite{mueller}  Of course, we 
are not able to compute the full $n$th order coefficients,
but it is not too difficult to identify subintegrals
that take the form 
\bea
C_0^{(0)}&\sim&(1/Q^4)\; \int_0^{Q^2} dk^2 k^2\; \ln^n(k^2/\mu^2) \nonumber \\
&\sim&
(1/2)\left(1/2\right)^n\; n!\, .
\label{con}
\eea
Diagrams 
that contribute to this result are illustrated 
in Fig.\ \ref{irree}, where $H$ absorbs lines
that are off-shell by order $Q^2$, and $T$ lines that 
have soft momenta.  Each power of the logarithm in Eq.\ \ref{con} is associated
with a self-energy (or other one-loop subdiagram that requires
renormalization).  A little renormalization group footwork shows that
the sum of all diagrams of this sort takes
the remarkably simple form
\bea
&& \int_0^{Q^2} dk^2 k^2\;  \alpha_s(k^2)  \nonumber \\
&& \quad = \int_0^{Q^2} dk^2 k^2\;  
{\alpha_s(Q^2) \over 1 + \left ({\alpha_s(Q^2)\over 4\pi}\right ) b_2\ln(k^2/Q^2)}\, ,
\nonumber\\
\label{irint}
\eea
in which expanding the denominator
 of the one-loop perturbative running coupling in
powers of $\alpha_s(Q^2)$ generates all the $n!$ coefficients.   
Behavior of this sort, which is associated with the (infrared) singularity
of the running coupling (at $k^2/Q^2=\exp[-4\pi/b_2\alpha_s(Q^2)]$), is known as an ``infrared renormalon". \cite{thooft}
If we still want to go to all orders, we would be tempted to
regularize perturbation theory by removing contributions from $k^2\sim \Lambda^2_{\rm QCD}$ in
all integrals like those in Eq.\ \ref{irint}, and replacing them by new,
nonperturbative parameters.  
This may be done in a surprisingly simple fashion. \cite{mueller}
 We can use gauge invariance arguments to
show that there is only one such parameter, which may be identified 
with the gluon condensate,
\be
C_{\rm PT}\rightarrow C_{\rm PT}^{\rm reg}+{\alpha_s \langle0|F^2(0)|0\rangle \over Q^4}\, .
\ee
Order-by-order in perturbation theory, the vacuum expectation value of the
gluon condensate generates all of the integrals of Eq.\ \ref{irint} above,
for which the upper limit appears as a renormalization scale.  Hence, to order
$1/Q^4$, it is natural to replace the factorial growth from soft gluons by
a nonperturbative value for the gluon condensate.
In summary, if there were no OPE, it would be necessary to invent it, to derive
useful information from infinite-order perturbation theory.

Could something similar be going on for resummed cross sections, leading not
to the OPE but to something else?  There are reasons to suggest that this might be
so, although to my knowledge no general formalism yet  exists. 
The simplest, but still very subtle, phenomenological  implications for the
nature of power corrections associated with infrared renormalons have been
widely discussed, especially in the context of $1/Q$ and/or $1/Q^2$ 
corrections in ${\rm e}^+{\rm e}^-$ event shapes \cite{shapeiir} and Drell-Yan cross sections. \cite{dyirr,korst}
For the former, the DELPHI collaboration has employed the recently-increased
LEP energy to compare theoretical expectations to data over a very
wide range of c.m.\ energies.   \cite{Duchen}

An example where some progress toward an operator formalism has been made is in 
Sudakov resummation for the transverse momentum distribution of a Drell-Yan pair,
noted above in Eq.\ \ref{list}. \cite{qtresum}  Here, as was shown years ago, the 
natural transformation (the moment of Sec.\ 4 above) is to impact parameter space, in terms of
which the transformed cross section includes a factor ${\tilde P}(b)$
that resums all logarithms of $b$, according to
\bea
\ln {\tilde P}(b)
&=&
\int_0^{Q^2}{dk_T^2\over k_T^2}A(\alpha_s(k_T^2))\ln(Q^2/k_T^2) \nonumber \\
&\ & \times \bigg ({\rm e}^{-ik_T\cdot b}-1 \bigg )\, ,
\label{CSexp}
\eea
where $Q$ is the mass of the produced pair, and where at leading logarithm $k_T$ 
may be interpreted as the transverse
momentum of a gluon.  In Eq.\ \ref{CSexp}, we have suppressed various
nonleading terms for simplicity. Up to such corrections, this
is  of the form of Eq.\ \ref{sudexp}
order-by-order in logarithms of $b$, once $b$ is identified as $iN/Q$.
In this form, however, as in Eq.\ \ref{sudexp}, the perturbative running coupling
passes through a region where it diverges.  Following our reasoning for
${\rm e}^+{\rm e}^-$ annihilation above, however,  we may  
replace that region by one or more nonperturbative parameter.  This leads to the
classic form suggested by Collins and Soper, \cite{qtresum}
\bea
\ln {\tilde P}(b)
&=&
\int_{(1/b^*)^2}^{Q^2}{dk_T^2\over k_T^2}A(\alpha_s(k_T^2))\ln(Q^2/k_T^2) \nonumber \\
&\ & \times \bigg ({\rm e}^{-ik_T\cdot b}-1 \bigg ) \nonumber \\
&\ & + g_1b^2\ln(Q^2b^2)+\dots\, .
\eea
Here $g_1$ is our nonperturbative parameter,
which replaces the integral $k_T<1/b^*$, 
\be
\int_0^{(1/b^*)^2}
{dk_T^2\over k_T^2}A(\alpha_s(k_T^2))\ln(Q^2/k_T^2)\big (b^2k_T^2\big )+\dots\, ,
\label{remove}
\ee
with $b^*$ an adjustable parameter
(upon which $g_1$ depends).

\begin{figure}
\centerline{\epsffile{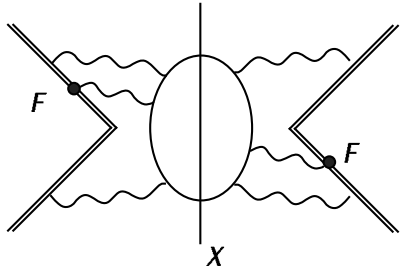}}
\caption{Representation of nonlocal vacuum expectation value.}
\label{Fsquare}
\end{figure}

We now ask whether there is an operator, whose 
perturbative vacuum expectation value
is identical to the infrared perturbative contribution \ref{remove}, which we have
removed?  In fact, there is, at least for $g_1$,
and it turns out to be a rather direct, if nonlocal,
generalization of the gluon condensate encountered in the OPE above. \cite{korst}
It is illustrated in Fig.\ \ref{Fsquare}, and consists of a pair of 
field strengths integrated over the paths of  two light-like eikonalized quarks,
like those encountered above in the resummation of threshold corrections for heavy 
quark production (Fig.\ \ref{HS}).   The eikonal lines, which may be
written as path-ordered exponentials of the gauge field,
\bea
&&\Phi_p[Sp+x,S_0p+x]  \nonumber \\
&&\quad ={\rm P} \exp \bigg ( -ig\int_{S_0}^S d s\, p^\mu A_\mu(ps+x) \bigg )\, ,\nonumber\\
\label{phioexp}
\eea
enforce gauge invariance in the nonlocal condensate, defined as
\bea
&&\langle 0| 
\Bigg|\Phi^\dagger_{p_2}(0,-\infty) \left( {\vec{\cal F}_{p_1}(0) - \vec{\cal F}_{p_2}^\dagger(0)}\right )\nonumber \\
&&\quad \quad \times
\Phi_{p_1}(0,-\infty)\Bigg |^2 |0\rangle\, ,
\label{tildEop2}
\eea
where ${\cal F}$ is constructed from $F^{\mu\nu}$ by
\bea
{\cal F}_{p}^\alpha(x)
&=& -ig\int_{-\infty}^0 d s\ \Phi_p(x,x+sp) \nonumber \\
&\ & \quad \times p_\mu F^{\mu\alpha}(sp+x)\; \Phi_{-p}(x+ps,x)\, .\nonumber \\
\label{Fdef}
\eea
Compared to the local gluon condensate of the OPE, the field strengths
are now free to migrate along a one-dimensional manifold of
lightlike paths, whose directions have been determined by the
incoming particles.  In this sense, it is a natural first
generalization of the OPE.   Correlations
in even higher dimensions are almost
certainly necessary, however, when dealing with
quantities like event shapes, which are sensitive to 
interactions at very large times.  It is my hope that
the clarification of these issues will be one of the interesting developments of the
coming years.

\section{Optimistic Summary}

I would like to suggest that the qualitative 
and quantitative successes of  perturbative QCD
this year 
are a prologue. 

Each of the major currents of research touched on in this
talk point to  further challenges.
The low-$x$ and diffractive behavior of DIS structure functions
as seen at HERA 
is suggesting new avenues to connect hadronic
and partonic degrees of freedom.  The BFKL program continues
to inspire inventive theoretical and experimental advances.
At the same time, Sudakov
resummation methods are
making it possible to understand the successes 
and limitations of NLO
calculations of high-$p_T$ and related cross sections at past and future
collider runs.  
There is a also a growing awareness of the need for 
a theory of power corrections beyond the OPE, hints
for which may be found in infrared renormalon analysis
of power corrections.  The
extended energy reach of LEP affords new opportunities
to  check and develop these ideas.  There were other related and important topics
discussed at the  conference
which space does not allow me to discuss here
beyond the merest mention, including the treatment of spin and nuclear effects
in hard scattering, photoproduction
and direct photon production, and elastic scattering.  
Many of the same methods, opportunities and challenges sketched above apply
to these topics as well.

Theorists will have to work hard to keep up with data  like those of
this year.  Now we are really ready to study  
quantum chromodynamics.

\section*{Acknowledgements}  I would like to thank the organizers of ICHEP96
for their invitation, and Wu-Ki Tung for helpful
conversations.  This work was supported in part 
by the National Science Foundation under
grant PHY930988.

\section*{Questions} 
{\it G.\ Farrar, Rutgers Univeristy:}

Will theory be able to  constrain power-law corrections,
e.g.\ those you mentioned for LEP event shape variables, or will they have to be fit as
for higher-twist corrections in DIS?

\vskip 12pt
\noindent{\it G.\ Sterman:} 

Parameters describing power-law corrections are
certainly nonperturbative, and must  be taken from experiment, like parton
distributions.  It is unlikely that a few numbers will describe these corrections.
It is my expectation, however, that a well-defined set of functions will turn
out to control power corrections to many event shapes. Because these corrections
are relatively large, they may well be more accessible experimentally than are higher-twist
corrections in DIS.

\vskip 12pt
\noindent{\it G.\ Wolf, DESY:}  

You have alluded to diffraction scattering in
BFKL and DGLAP.  You seem to
suggest that diffraction scattering should be stronger in
BFKL as compared to DGLAP.  Did I  understand you correctly?

\vskip 12pt
\noindent{\it G.\ Sterman:} 

Yes, although I would put it
slightly differently.  I would suggest that the diffractive component
of DIS structure functions evolves by a  process that is of the BFKL type,
lacking strong ordering in transverse momenta.

\end{document}